\documentclass[english,prl,showpacs,amsmath,amssymb,aps,mathbbold,12pt]{revtex4-1}

\usepackage[latin9]{inputenc}
\usepackage[english]{babel}
\usepackage{esint}
\usepackage{braket}

\usepackage{graphicx}
\usepackage{epstopdf} 
\usepackage{pdfsync}
\usepackage[dvipsnames]{xcolor}
\usepackage[normalem]{ulem}
\usepackage{hyperref}
\usepackage{slashed}
\usepackage{dsfont}
\usepackage{float}
\usepackage{comment}

\newcommand{\be}{\begin{equation}}   
\newcommand{\ee}{\end{equation}}

\newcommand\half{\frac 1 2 }

\newcommand\oncite [1] {Ref.\, \onlinecite{#1} }
\newcommand\oncites [1] {Refs.\, \onlinecite{#1} }

\begin{document}
\title{ Quantum Hall Liquids Coupled to Dynamical Electromagnetism}

\author{T. H. Hansson}
\affiliation{Department of Physics, Stockholm University, Sweden} 
\author{Qing-Dong Jiang}
\affiliation{Tsung-Dao Lee Institute \& School of Physics and Astronomy, Shanghai Jiao Tong University, Shanghai, 201210, China}
\author{S. A. Kivelson}
\affiliation{Dept.\ of Physics, Stanford University, Stanford, CA., USA }
\author{Thomas Klein Kvorning}
\affiliation{Royal Institute of Technology, Stockholm, Sweden}

\begin{abstract}
 We investigate the effect on a Quantum Hall (QH) liquid of its coupling to 3+1 dimensional dynamical electromagnetism, which renders the system gapless.  We calculate both the Hall and longitudinal resistances, $\rho_H$ and $\rho_L$, in the context of a minimal model of the electromagnetic environment, with a small three dimensional conductivity $\tilde\sigma$, that allows for a counter-flow current.  In the thermodynamic limit, we show that  $\rho_H$ is quantized, while $\rho_L$ approaches a non-zero limit, $\rho_L \sim \alpha\, R_K$, where $\alpha$ and $R_K=2\pi /e^2$ are the fine structure and the Klitzing constant. In contrast, the QH conductance, $\sigma_H$, is smaller than the expected quantized value by a correction $\sim \alpha^2/R_K$. 
The electromagnetic interaction also generates corrections of order $\alpha^2$ to the quasiparticle charges and statistics, in a way that is consistent with general arguments based on gauge invariance.
In addition, we present an intuitive argument that relates the flux attachment associated with the composite boson representation of the electron liquid to the empirically observed 
approximate quantization of $\rho_H$, even in circumstances in which $\rho_L$, and the deviation of $\sigma_H$ from its quantized value, are substantial.
\end{abstract}
                                        
\maketitle                 

 \emph{Introduction} -- 
The integer quantum Hall (QH) effect is characterized, in the $T \to 0$ limit, by vanishing longitudinal resistivity and quantized Hall resistivity: $\rho_L \to 0$ and $\rho_H \to  R_K/\nu$, where $R_K \equiv h/e^2$ is the von Klitzing constant and $\nu$ is an integer labeling distinct phases. The quantization of $\rho_H$ is so precise that it is used meteorologically, with relative deviations below $ O(10^{-9})$~\cite{delahaye2002ac,ahlers2009compendium}. This precise quantization was originally explained by Laughlin \cite{laughlin1981} and Halperin \cite{halperin1982} using arguments based on gauge invariance.
A different theoretical perspective was later introduced by Thouless et al.\cite{tknn,ntw}, who used the Kubo formula to show that the Hall conductivity, $\sigma_H$, of any non-degenerate ground state is proportional to the first Chern number, which is an integer-valued topological invariant characterizing the electronic state. 
The Thouless analysis assumes the existence of a gap, and the Laughlin/Halperin 
argument a mobility gap.  Later, it was understood how some of the arguments related to the integer QHE generalize to the observed fractional QH states, and that theories for these based on mean field theories of composite bosons or fermions, and related hydrodynamics, could successfully predict many patterns and properties of the fractional QH liquids. However, in actual solid state QH experiments the electrons are coupled to gapless photons and phonons. Aspects of this 
have been discussed, for instance in \oncites{haldane1984magnetic,PhysRevB.87.195451}, but we know of no systematic investigation of topological properties. Here, we will explore the effect of the coupling between the 2d electronic system (that supports the QH state) and the quantum fluctuations of the electromagnetic field in the surrounding (3d) space. There are related recent studies of the effects of cavity modes which are consistent with the results in this paper \cite{cardoso2026cavity,yang2026quantum}.  (See also Sect. III in the Supplementary Material (SM)).

It is clear that with such a coupling a uniform Hall current at frequency $\omega$, will dissipate energy due to radiative loss. What is less obvious, although well known, is that in the 
{appropriately defined} $\omega\rightarrow 0$ limit, $\rho_L\rightarrow \ Z_0/2$ where $Z_0= 2\alpha R_K \approx 377\, \Omega$ is the ``impedance of the vacuum'' and  $\alpha=e^2/hc\approx 1/137$ is the fine-structure constant. Consequently, $\sigma_H$ and $\rho_H$ cannot both be precisely quantized. The central result of this paper is that the robustly quantized quantity in the presence of electromagnetism is $\rho_H$ rather than $\sigma_H$.
More generally, in the discussion section, we interpret the remarkable robustness of the quantization of $\rho_H$ as a physical manifestation of the linking of charge and flux currents in the composite boson representation of the quantum Hall fluid.  In this context, we reference a number of experimental situations in which robust quantization of $\rho_H$ is seen in systems with substantial values of $\rho_L/R_K$. 

We proceed using a simplified model of a QH liquid coupled to dynamical electromagnetism. We consider an infinite plane that hosts a 2d electron fluid and a homogeneous positive neutralizing background charge density, which is encased by thin perfectly insulating layers and embedded in an infinite 3d space filled with a weakly conducting medium. 
The conductivity, $\tilde \sigma$, of the 3d space is included to regularize the problem; it allows for a current counter-flow, without which there would be an infinite magnetic energy associated with a uniform in-plane DC current. Typically, we will emphasize results in the limit $\tilde\sigma\to 0$ once physical quantities have been evaluated. Our analysis establishes as an important point of principle the existence of a case in which $\rho_{xy}$ is perfectly quantized while $\rho_{xx}$ is finite.  It is important to note, however, that the result is obtained in the thermodynamic limit - a limit that is not, actually, relevant in any real experiment as discussed in the concluding section.
(However, we will also propose an  experimental geometry in which the radiative loss can be  tuned, allowing for a direct test of its effect on the quantization of $\rho_H$.)

We use the Wen-Zee hydrodynamic theory \cite{wenzee1992}, to model the QH fluid, since it provides a general description in the topological scaling limit. Although this theory does not depend on any specific microscopic model or wavefunction, it can be derived using a saddle-point approximation to an exact microscopic model expressed in terms of composite bosons \cite{zhk,zhangrev}. We have also used 
this Ginzburg-Landau-Chern-Simons (GLCS) theory to directly derive our results.

\emph{The model} --
We model the QH liquid using a hydrodynamic gauge field $b_\mu$, and the fluctuating electromagnetic vector potential $Q_\mu$ with a Maxwell action with permeability  $\mu=1$ and dielectric function, $\epsilon(\vec p,\omega)$ of the embedding 3d space.  We also introduce a (classical) externally applied  field, $A_\mu$ for later convenience.

For the QH fluid at {the Laughlin}  filling fractions $\nu = 1/k$ in the $(x,y)$ plane, we take the action, 
\begin{align} 
S_b = \int d^3x\, &\left[ -\frac k {4\pi 
}\epsilon^{\mu\nu\sigma}   b_\mu  \partial_\nu b_ \sigma   +  \frac e {2\pi c} \epsilon^{\mu\nu\sigma} \left ( 
A_\mu + Q_\mu \right) \partial_\nu b_\sigma + j_v^\mu b_\mu  \right. \nonumber \\
& \left.+ \frac m {2 \bar\rho} E_b^2 - \half V''(\delta\rho^2)  +\frac 1 {8m\bar\rho}\vec\nabla\rho\cdot\vec\nabla\rho   \right]  \, . 
\label{Sb}
\end{align}
where we have 
included a factor of $e/c$ 
in the coupling to the electromagnetic fields that is often omitted.

The first line of Eq. \ref{Sb} is the Wen-Zee hydrodynamic theory \cite{wenzee1992}, 
while the second line contains higher order terms with coefficients that can be derived from the 
Ginzburg-Landau Chern-Simons (GLCS) 
effective action - i.e. the representation of the 2DEG in terms of composite bosons \cite{zhk}.  $V[\rho]$ is a potential that is minimized at $\rho = \bar\rho=\nu eB/2\pi c$, where $B$ is a constant 3d background magnetic field in the $z$-direction, and $\rho = \bar\rho + \delta\rho$. The hydrodynamic gauge field, $b_\mu$ is normalized so that the electromagnetic charge and current densities are
$\rho = \frac e {2\pi c} \epsilon^{ij}\partial_i b_j=\frac e {2\pi c} B_b$ and $J^i = -\frac e {2\pi }\epsilon^{ij}(\partial_t b_j-\partial_i b_0) =\frac e {2\pi } E_b$ where up to a factor $k$, $B_b$ and $E_b$ are the statistical and magnetic fields of the GLCS theory. Also, $m$ is the electron effective mass, $j^\mu_v$ is the quasiparticle current (proportional to the vortex current, hence the subscript $v$), and $\bar \rho$ is matched by a compensating  uniform background charge density of opposite sign.

For the dynamical transverse field $\vec Q$  we take  a Maxwell action with  permeability $\mu=1$, and a frequency and momentum dependent dielectric function 
\begin{equation} \label{diel}
\varepsilon(\vec p,\omega )  =  \varepsilon_0  - \frac {\tilde\sigma } {i\omega - Dp^2}\, ,
\end{equation}
which interpolates between  Thomas-Fermi screening of the Coulomb interaction at $\omega=0$:
$\varepsilon(\vec p,0)  = \epsilon_0 - (\kappa/p)^2$,
with $\kappa^2 = \tilde\sigma/D$, and Drude behavior for $\vec k=\vec 0$:  
$\varepsilon (\omega)= \varepsilon_0 + i\tilde\sigma  /\omega \, ,$ which describes the (imaginary) dielectric response of a 3d conductor with conductivity $\tilde\sigma$. 
%
%
(Note that, in natural units, the 3d conductivity, $\tilde\sigma$, has units of frequency while the 2d quantities, $\sigma_L$ and $\sigma_H$, have units of velocity.) The expression for $\varepsilon (\vec p,\omega)$ is valid at asymptotically low frequencies and momenta, where $|\vec p|$ and $|\omega |$ are much smaller than any other static or dynamic scale of the medium. 
We will always assume that $\tilde\sigma$ is 
small but non-zero, although in computing quantities for which the limit is non-singular, we will sometimes take $\tilde \sigma \to 0$.

We will use the Coloumb gauge, $\vec\nabla\cdot\vec Q=0$, and write the Maxwell action coupled to the above (2+1) dimensional sources as,  
\begin{align} \label{genact}
    {\cal S}_{M}=\frac 1 2 \int \frac{d^4p} {(2\pi)^4} 
    \left\{\vec Q(-p) \left[(\omega/c)^2\varepsilon(\vec p,\omega)-q^2 -p_z^2  \right]\cdot \vec Q (p) +
 Q_0(-p)\, \varepsilon(\vec  p,\omega)p^2 Q_0 (p) \right\} 
\end{align}
where we have used the notation  $p =(\vec p, \omega) ={(
\vec q},p_z,\omega)$, and  $q=|\vec q|$, with $\vec q$  a two-vector  in the plane of the QH liquid.  Currents and charges confined to the $(x,y)$ plane have no $p_z$ dependence so adding the source term $J^\mu Q_\mu$ to \eqref{genact} and integrating $Q_\mu$, we get an $O(\alpha)$ correction term to the hydrodynamic action
$S_b \to S_b + S_{em}$, with
\begin{equation} 
S_{em}[b_\mu] = 
\half \left(\frac {e} {2\pi }\right)^2 \int \frac{d^4p}{(2\pi)^4} \left[\vec b(-p)\,\frac {-\omega^2 } {\omega^2\varepsilon(\vec p,\omega)-c^2q^2 -c^2p_z^2 + i\epsilon   }\cdot \vec b(p) +  \rho(-p)\frac 1 { \varepsilon(\vec  p,\omega)p^2} \rho (p)  \right] \, . 
\label{Sem}
\end{equation}
So far we have not made any approximations, but you should keep in mind that the 
action in Eq. \ref{Sb} itself follows from the microscopic composite boson formulation of the problem only to quadratic order in fluctuations about a saddle-point (mean-field) configuration.

The second term in Eq. \ref{Sem}       primarily affects the properties of static or quasi-static charge-density variations.  It encodes the Thomas-Fermi screening  in the limit  $|\omega| \ll 
{D |\vec q|^2}$, so is important, for instance, in considering the properties of any charges that are localized in the presence of disorder.
Since for understanding transport we are typically interested in the opposite order of limits (i.e. $|\vec q|$ small compared to the appropriate power of $|\omega|$), we will henceforth neglect this term and focus on the first.

To proceed, we must carry out the $p_z$ integration in the first term in 
Eq. \ref{Sem}, and although this can be done analytically, the resulting expression is unwieldy, and hard to analyze. However, in the  
SM we show that in the parameter range of interest, one can neglect the $p_z$ dependence in the dielectric function and then carry out the $p_z$ integration in Eq. \ref{Sem} to get the corresponding contribution to the effective 2+1 dimensional action,
\begin{equation} \label{pzres}
S_{em}[\vec b] = \frac 1 4 \left(\frac {e} {2\pi }\right)^2 \frac 1 c\int \frac{d\omega d^2q}{(2\pi)^3}\vec b(-\omega,-\vec q)\,\frac {i \omega^2 } {\sqrt{\omega^2 + i\tilde\sigma \omega -c^2q^2 }}\cdot \vec b(\omega,\vec q) \, ,
\end{equation}
where we have set $\epsilon_0 =1$ \cite{footnote1}.

\emph{EM corrections to the Quantum Hall resistance } --
To calculate the Hall resistances we put $\vec q = \vec 0$ and we can also neglect higher derivative terms, as well as the terms $\sim\vec\nabla\rho$ in \eqref{Sb}. 
Under conditions such that the quasiparticle current vanishes, the total bosonic action $S_{tot} = S_b + S_{em}$ in momentum space becomes
\begin{equation} \label{baction}
    S_{tot} =   \int \frac{d^3p} {(2\pi)^3} \left[ \frac\alpha {4\pi} \frac {i \omega^2 } {\sqrt{\omega^2 + i\tilde\sigma \omega  }} \delta^{ij}b_i b_j - \frac k {4\pi }i\omega\epsilon^{ij}  b_i b_j-\frac e {2\pi} \epsilon^{ij} b_i E_j \right] \, , 
\end{equation}
where $E_i = \frac {i\omega} c A_i$. Varying with respect to $b_i$ gives the equation of motion,
\begin{equation} \label{eom}
    \frac \alpha {2\pi}     \frac {i \omega^2 } {\sqrt{\omega^2 + i\tilde\sigma \omega  }}b^i - \frac k {2\pi c}i\omega\epsilon^{ij}  b_j -\frac e {2\pi}  \epsilon^{ij} E_j=0 \, ,
\end{equation}
where  we note that there is no gap. Using $ J^i = \frac e {2\pi } i\omega \epsilon^{ij} b_i$, this becomes
\begin{equation} \label{eom2}
    -\frac \alpha e 
    \frac { \omega } {\sqrt{\omega^2 + i\tilde\sigma \omega  }}\delta^{ij}J_j -\frac k {e } \epsilon^{ij}J_j +\frac e {2\pi}  E^i=0
\end{equation}
from which we obtain the 2d transverse and longitudinal resistivities, defined by $E^i = (\rho_L \delta^{ij} + \rho_H \epsilon^{ij} )J_j$: 
\begin{equation} \label{results}
   \rho_H = k \frac{2\pi}{e^2} = k R_K \quad\quad ;\quad\quad \rho_L =    \frac { \omega } {\sqrt{\omega^2 + i\tilde\sigma \omega  }} \alpha R_k  =    \half\frac { \omega } {\sqrt{\omega^2 + i\tilde\sigma \omega  }} Z_0\, .
\end{equation}
Thus, the Hall resistivity is not affected by the electromagnetic interactions, but the longitudinal (and hence also $\sigma_H$) are shifted from their values in an isolated Hall liquid. To obtain the essential, medium independent result, we first take the limit $\tilde \sigma \to 0$ and then $\omega\to 0$.
In this limit $\rho_L \rightarrow \half Z_0$. A direct calculation shows that this reflects the power carried by the emission of electromagnetic radiation. (See Sect III of the SM for details.)

The above derivation can straightforwardly be extended to the case of several Landau levels and hierarchy states by introducing several hydrodynamical fields. 


\emph{QH conductance, fractional charge and  statistics} --
To determine the electromagnetic effects on the fractional charge and statistics of the quasiparticles, we derive the pertinent response action $S[A,j]$,
\begin{align}
{\cal S}_{QH}(A,j,Q,b) &=\half \int\frac{d^3p }{(2\pi)^3} \, \left(b_\mu K(\omega,\vec q)b^\mu- \frac k {2\pi }  \epsilon^{\mu\nu\lambda} b_\mu ip_\nu b_\lambda 
- \frac e {2\pi} \epsilon^{\mu\nu\lambda} (ip_\nu A_\lambda)b_\mu  - j^\mu b_\mu  \right) \, .
\end{align}

The kernel $K(\omega, \vec q)$ and the corresponding Green function are given in the 
{SM}, where we also give the details of how to integrate the field $b_\mu$ from  which we extract the conductivity tensor, $\bar\sigma$ as well as as well as the fractional charge $e^\star$ and statistical angle $\theta_s$ for the anyonic quasiparticles. As expected, $\bar\sigma = \bar\rho^{-1}$ where $\sigma_H = 
f(x)/\rho_H$, where 
\begin{equation} \label{hallcond}
f(x) =\left( 1 + x^2\right)^{-1} \approx  1- x^2 \ , \  \ x =\rho_L/\rho_H \, .
\end{equation}
Importantly, the fractional charge, \cite{footnote2}
and statistics are corrected by the same factor as $\sigma_H$, i.e. $e^\star = 
f(x)\frac e k$ and statistical angle $\theta_s =  f(x)\frac \pi k$. 
{In Sect. IV of the SM, we rederive these results using the  GLCS theory. The generalization to the hierarchy states is less direct, but might be done using the methods in \oncite{tournois2020microscopic}. }


\emph{Intuitive explanation based on Composite Bosons} --
The electric current can be written either as $J^i = \frac e {2\pi} \epsilon^{ij} b_j $ or as $J^i = \frac e {2\pi k} \epsilon^{ij} a_j = \frac 1 k \tilde J^i $ (where $a_\mu$ is the statistical gauge field in the composite boson formulation), so, in the bosonic description, up to the scale factor $1/k$, the electric current \emph{is} the CB current $\tilde J^i$. Thus the second term in \eqref{eom2}, which gives the QH response, directly reflects that the current carriers are charged bosons and the Hall voltage is a direct consequence of their moving flux \cite{scientificamer}. The first term, which only contributes to the $\rho_L$ is due to the coupling to the 3d e.m. field. We might imagine other contributions to this term (e.g. higher order corrections in powers of $\alpha$), but as long as there are no other parity violating effect except for the external $B$ field, it will not effect $\rho_H$. In fact, it is widely observed -- although not always remarked upon --  in various experimental situations involving either more strongly disordered samples or slightly elevated temperatures, that $\rho_{xy}$ exhibits well developed QH plateaus although $\rho_{xx}$ is a substantial fraction of $R_K$ \cite{serlin2020intrinsic,longju,checkelsky,xiaodong}.  Indeed, sometimes even in an ``insulating'' regime in which $\rho_{xx}$ is both $ \gg R_K$ and is a strongly increasing function of decreasing $T$, a roughly $B$ and $T$ independent $\rho_{xy}$ has been observed with a value that is close to the quantized value $\rho_{xy}=1 \ R_K$ \cite{shahar}.  Thus, there is empirical evidence that the (at least approximate) quantization of $\rho_{xy}$ is considerably more robust in actual experimental circumstances than  might be expected on general grounds \cite{footnote3}

We now outline a more formal way of describing CB transport, which is an extension of the analysis  in Refs. \cite{global,scientificamer}. 
 The CBs respond to the sum of the electric and the statistical electric fields, so assuming the fluctuations of the statistical electric field about its mean are small enough that  linear response theory applies, we have{
$\varepsilon_i - e E_i = \rho^{(cb)}_{ij} \tilde J^i$
    where $\rho^{(cb)}_{ij}$ is the composite boson resistivity tensor
    and $\vec \varepsilon_i = \partial_i a_0-\partial_0 a_i$ is the statistical electric field.
In the condensed phase, and in the absence of long-range interactions, we expect} that  $\rho_{ij}^{cb}=0$, but it is non-trivial to compute in a non-condensed phase. None-the-less, proceeding formally, we have
\begin{equation}
    E^i = \frac 1 e (2\pi k \epsilon_{ij} - \rho^{(cb)}_{ij}) \tilde J^i =\rho_{ij} J^j
\end{equation}
so 
\begin{equation}
    \rho_{ij} = k \frac {2\pi} {e^2} \epsilon_{ij} - \frac 1 {e^2} \rho^{(cb)}_{ij} \,.
\end{equation}
At 
saddle point level, the composite bosons
feel an effective magnetic field
$
    B^{eff}(\vec r) = B - \frac k {2\pi} \tilde J^0(\vec r)\, ,
$
which, in the presence of disorder or some form of charge-density-wave order may be a function of position).  However, for $B$ near the commensurate value, $B= B_k \equiv k/(2\pi) \bar J^0 $, the field that the composite bosons see is small - suggestive that $\rho_{xy}^{(cb)}$ is generically small.  More formally, in an uncondensed phase,  $\rho_{xy}^{(cb)}$ must be a continuous function of $B$, which we expect will change sign from the regime where $B \gg B_k$ to $B\ll B_k$.  It therefore must pass through 0 for a value of $B^\star \approx B_k$.  It follows that as a function of $B$, independent of the magnitude of $\rho_{xx}$, so long as the composite boson picture applies, $\rho_{xy}$ must take the value $k R_K$ for a field $B^\star \approx B_k$.  Moreover, so long as the further intuition based on the smallness of $B^{eff}$ applies, this should be the center of an approximate plateau. In this sense, the approximate quantization of $\rho_{xy}$ confirms the physical import of flux attachment \cite{scientificamer}, even when the CBs are not condensed \cite{footnote4}.


\emph{Discussion and summary} -- A realistic model for the electromagnetic environment in a typical QH experiment would be quite complicated, but it is easy to estimate the finite size effect. 
 The momentum $\vec q$ in \eqref{pzres} will be cut off at $q \approx 2\pi/L$, where $L$ is the size of the Hall bar.  With  $\omega = 10\,{\rm Hz}$ and $L = 2\times 10^{-4}$ m, the $\rho_L$ at this cutoff value, below which our calculation cannot be trusted, is $\rho_L \sim (\omega/c q ) Z_0\approx 10^{-12} Z_0 $, which is well below the experimental lower limits \cite{delahaye2002ac,ahlers2009compendium}.  However, if we consider another simplified but plausible experimental geometry, the effect 
 {could} be sufficiently amplified  to be measurable.  Consider a Hall bar of large size $L$, again encased in insulating layers that confine the current flow in the 2DEG, but placed inside a symmetric pair of capacitor plates a distance $d \ll L$ from the 2DEG.  The electro-magnetic coupling to the screening gates then produces a backflow current to confine the induced magnetic field within the device. 
 As long as radiative loss from the edges is small, this leads to a dissipation corresponding to a  $\rho_L = (1/2)\rho_{gate}$  where $\rho_{gate}$ is the 2d resistivity (in $\Omega/\mathrm{Area}$) of the gates. 

As we have shown above, the same correction factor $f(x)$ appears in all the three 
quantities, $\sigma_H$, $e^\star$ and $\theta_s$, and this is highly significant. Laughlin constructed the ``Laughlin quasihole'' by considering 
the insertion of a unit flux into a QH liquid, and from this it is clear that if the Hall conductivity and the fractional charge are renormalized, it must be by the same factor.
Also, in  \oncites{kivelson1985,Hansson_2025} it was shown, again using gauge invariance, that in a quantum Hall system with Hall conductivity $\sigma_H = \frac 1 k \frac {e^2} {2\pi}$ and fractional charge $e^\star = e/k$, the fractional statistics must be $\theta_s  = \pi/k$. In these papers, $k$ was taken as an integer, but the argument is valid for any value of $k$. It is thus a matter of consistency that the modified values of $\sigma_H$, $e^\star$ and $\theta_s$ take the forms presented in the SM (Sec. III).

What is measured in experiments are resistance, not conductance, but it is commonly assumed that also $\sigma_H$ is quantized. As mentioned in the introduction, the theoretical argument for this is the relation to the first Chen number, and experimentally it is strongly supported by the measured $\rho_L$ which is zero within experimental limits, at least at the center of the integer plateau.
The theoretical argument for the quantization of the Hall resistance is based on gauge invariance, and the results in this paper indicate that this is the more fundamental insight, in that it is robust against coupling to dynamical electromagnetism. 

Finally, we would like to again stress the main message of this paper: Although there is per se nothing surprising with a radiative loss at finite $\omega$, that the effect remains in the $\omega\rightarrow 0$ is somewhat surprising. The main surprise, however, is that the Hall resistance $\rho_H$ remains at its quantized value, with no electromagnetic corrections, in spite of the system being gapless. We believe that this unexpected robustness might have a significance that even goes beyond the arguments based on gauge invariance.  

\noindent
{\bf Acknowledgments} \\
We thank Steve Simon, Shivaji Sondhi and Ady Stern for discussions. {This work was supported in part by NSF-BSF award DMR2310312 at Stanford. Q.D.J. acknowledges the support from National Natural Science Foundation of China under Grant No. 12374332, Cultivation Project of Shanghai Research Center for Quantum Sciences Grant No. LZPY2024, and Shanghai Science and Technology Innovation Action Plan Grant No. 24LZ1400800. The research of T.K.K. is funded by the Wenner-Gren Foundations.}


\begin{thebibliography}{}
\bibitem{footnote1}
A similar expression - but in a different context - was derived in Ref. \cite{fradkin}.
\bibitem{footnote2}
Our analysis of the quasi-particle charge neglects the effects of Thomas-Fermi screening, reflected in the heretofore neglected second term in Eq. 4.  This is valid at distance scales short compared to the Thomas-Fermi screening length, $\kappa^{-1}$, of the 3d embedding space which diverges as its compressibility  tends to zero. 
Also, in the limit $q\ll\omega$, relevant for the statistics, the screening is an  $O( q^2/\omega^2)$ effect.
\bibitem{footnote3}
We are aware of three physically distinct theoretical analyses that have been investigated in the past that could underpin this observation:  1)  In an analysis of a 1D quantum Hall channel employing a Landauer approach, it was shown \cite{jainandme} that elastic backscattering  can give rise to a finite $\rho_{xx}$ when to the same level of approximation $\rho_{xy}$ is perfectly quantized.  This sort of mesoscopic physics is unlikely to be relevant to the large devices in which robust approximate quantization of $\rho_{xy}$ has been observed.  2)  In attempting to account for the observations in Ref. \cite{shahar}, it was noted that the ``semicircle law'' of classical percolation \cite{dykhne} in a system of quantum Hall and insulating puddles  results \cite{auerbach1} in quantized Hall insulating behavior.  Essential to this theory is that there is some form of long-wave-length disorder that results in puddles of coexisting phases on length scales long compared to any scale of the quantum coherent motion of charges.  It seems unlikely (but not impossible) that this applies in the various experimental systems in which such behavior is observed.  3) The same line of composite-boson-based reasoning \cite{global} that is expanded on in the main text initially led to the proposed existence of a ``Hall insulator'' phase proximate to a quantum Hall plateau with $\rho_{xy}$ close to its quantized value.
\bibitem{footnote4}
Note that since we are dealing with an uncondensed phase of the composite bosons, it may be useful to look at this from the dual perspective \cite{dunghai}, i.e. treating the vortices and anti-vortices as the dynamical degrees of freedom.  Here $\sigma_{Ij}^{(vort)}= \rho_{ij}^{(cb)}$, and $B^*$ is the point at which the contribution of vortex and anti-vortex currents exactly cancel. A somewhat similar  argument was presented in Ref. \cite{sri} in terms of composite fermions in favor of an emergent particle-hole symmetry  in the quantum Hall metal phase near $\nu=1/2$ at a field in the ``middle'' of the phase - analogous to our $B^*$.
\end{thebibliography}

\newpage
\section{Supplementary Material}
\noindent
{\bf I. The $p_z$ integral in Eq. \eqref{Sem}} 

Here we analyze the $p_z$ integral in the first term of  \eqref{Sem}.  First we notice that neglecting the $p_z$ dependence in $\varepsilon$, the $p_z$ integral is elementary and gives the result \eqref{pzres} in the main text. We now argue that, in the relevant parameter range, this is a valid approximation.

Clearly, the $p_z$ dependence can be ignored whenever $q \gg p_z$, so we take the worst case situation and set $q=0$. 
Then in the first term in Eq. 4,  the only important values of $p_z$ are $c^2|p_z| \sim \epsilon(\omega) \ \omega^2$ where $\epsilon(\omega)=\epsilon_0 - i\tilde \sigma/\omega$. Now there are two cases:\\
i) For $\tilde \sigma > \omega$ the important values of $p_z^2 \sim \tilde \sigma \omega/c^2$.  Now self-consistency, plugging  this into the denominator of Eq. 2, we find that it changes the value of $\epsilon$ by a fraction proportional to $Dp_z^2/\omega \sim  D \tilde \sigma/c^2$.  Note that because $\tilde \sigma$ is a frequency $D\tilde \sigma$ does indeed have units of velocity squared so this is a dimensionless factor.  It is small in proportion to the fine structure constant squared and also in proportion to $\tilde \sigma/\tilde \sigma_Q$ where $\tilde \sigma_Q$ is an appropriate 3d quantum of conductance - probably roughly the Mott-Ioffe-Regel values.  So this is indeed a small factor. \\
ii) For  $\tilde \sigma < \omega$ the important values of $p_z^2 \sim  \epsilon_0\omega^2/c^2$.  Now self-consistency, plugging  this into the denominator of Eq. 2, we find that it changes the value of $\epsilon$ by a fraction proportional to $\epsilon_0 D\omega/c^2$.  This is still small in proportion to the fine-structure constant squared, and also to the frequency, which although large compared to $\tilde \sigma$ is still supposed to be small compared to all electronic frequencies (that enter through $D$).

\noindent
{\bf II. Radiative loss and dissipation}

We consider a uniform oscillating current sheet located at $z=0$,
$
\vec J = J e^{-i\omega t}\,\hat x \, .
$
By symmetry, the electromagnetic fields take the form
$
\vec E = E_x(z)\,\hat x, ~ \vec B = B_y(z)\,\hat y \, .
$

In rationalized Gaussian (Heaviside-Lorentz) units, Maxwell equations read
\begin{align}
\label{amperelaw}\nabla \times \vec B &= \frac{1}{c}\partial_t \vec E + \frac{1}{c}\vec J, \\
\nabla \times \vec E &= -\frac{1}{c}\partial_t \vec B \, .
\end{align}
Assuming a local (momentum-independent) dielectric response
$
\epsilon(\omega) = 1 + \frac{ i\tilde\sigma}{\omega},
$
where $\tilde\sigma$ is 3d conductivity. 
The fields away from the sheet ($z\neq 0$) satisfy
\begin{equation}
-\partial_z^2 B_y = \frac{\omega^2}{c^2}\epsilon(\omega)\, B_y.
\end{equation}
Assuming solution of the form
$
B_y(z,t) = B_0 e^{-i\omega t - \kappa |z|}, ~ E_x(z,t) = E_0 e^{-i\omega t - \kappa |z|},
$
one obtains the dispersion relation
\begin{equation}
\kappa^2 = -\frac{\omega^2}{c^2}\epsilon(\omega).
\end{equation}
And according to Faraday's induction law, one further obtains the relation between $E_0$ and $B_0$, i.e.,
\begin{equation}
\kappa E_0 = -\frac{i\omega}{c} B_0 \, .
\end{equation}

Integrating Ampere's law Eq.\eqref{amperelaw} across the sheet, one obtains
$
B_y(0^+) - B_y(0^-) = \frac{1}{c}J.
$
Using symmetry $B_y(0^\pm)=\pm B_0$, we obtain 
\begin{equation}
B_0 = \frac{1}{2c} J \, .
\end{equation}

The Poynting vector is
\begin{equation}
\vec S = {c}\,\mathrm{Re}\left(\vec E \times \vec B^*\right).
\end{equation}
The total power radiated per unit area (both directions) is
\begin{equation}
\frac{P}{A} = 2 S(0) 
= 2c\,\mathrm{Re}(E_0 B_0^*)= \frac{1}{2c^2}\,
\mathrm{Re}\!\left(-\frac{i\omega}{\kappa}\right) J^2.
\end{equation}
Defining the radiative impedance via $P/A = \rho_{\rm rad} J^2$, we obtain
\begin{equation}
\rho_{\rm rad}
= \frac{ 1}{2c^2}\,
\mathrm{Re}\!\left(-\frac{i\omega}{\kappa}\right).
\end{equation}

Substituting $\kappa = \frac{1}{c}\sqrt{-\omega^2 - i \tilde\sigma\omega}$ gives the final form
\begin{equation}
\rho_{\rm rad}
=
\frac{1}{2c}\,
\mathrm{Re}\!\left(
\frac{\omega}{\sqrt{\omega^2 + i \tilde\sigma\omega}}
\right).
\end{equation}
In the vacuum limit $\tilde\sigma \to 0$, this reduces to
\begin{equation}
\rho_{\rm rad} = \frac{1}{2c} = \frac{Z_0}{2},
\end{equation}
where $Z_0 = 1/c$ is the vacuum impedance in rationalized Gaussian units. 

As a consistency check, let's also calculate the counter flow current, $J_{cf}$, generated in the ambient space
\begin{equation}
    J_{cf} = 2\int_0^\infty dz\, J(z) = 2 \tilde\sigma \int_0^\infty dz\, E_x(z) ={\tilde\sigma J \times \frac{i\omega}{\kappa^2c^2}\int_0^\infty d(\kappa z) e^{-\kappa z}}= -\frac {i\tilde\sigma} {\omega+i \tilde\sigma} {J} \, .
\end{equation}
For $\tilde\sigma = 0$ there is clearly no current generated, but just an undamped electromagnetic wave. For $\omega = 0$, we get $J_{cf} = -J$, which is needed not to have a constant magnetic field at infinity.


\noindent
{\bf III. QH conductance, fractional charge and  statistics} 


To find out how the dynamical electromagnetism influences the fractional charge and statistics of the quasiparticles, we derive the pertinent response action $S[A,j]$. We start from \eqref{baction} written as,
\begin{equation}
   {\cal S}_{QH}(A,j,b)=  \half \int\frac{d^3p }{(2\pi)^3} \,\left( b_\mu (G^{-1})^{\mu\nu}b_\nu  \frac e {2\pi} -\epsilon^{\mu\nu\lambda} (ip_\nu A_\lambda)b_\mu  - j^\mu b_\mu \right)  \, , \nonumber
\end{equation}
from which we derive the propagator,
\begin{equation} \label{propagator}
    G(p)_{\mu\nu} = \frac 1 {K^2 - L^2p^2} \left( K g_{\mu\nu}+ iL \epsilon_{\mu\beta\nu} p^\beta \right) \, ,
\end{equation}
where we have put $c=1$, $p =(\omega,\vec q)$, $L=k/2\pi$, and
\begin{equation}
    K(p) = \frac i 2 \left( \frac e {2\pi}\right) ^2 \frac {p^2} {(\omega^2 +i\tilde\sigma\omega -q^2)^\half} \, .
\end{equation} 

Let us first check that we reproduce the previous results for the resistivity. For this we ignore $j^\mu$, and put $\vec q = 0$, and use \eqref{results}, to get $K(\omega )= {-i\omega} \left(\frac e {2\pi}\right)^2 \rho_L$. This implies
\begin{equation}
    {K^2 - L^2p^2} = -\omega^2 \left(\frac e {2\pi}\right)^4[\rho_L^2 +\rho_H^2]
\end{equation}
which gives (for $A_0 =0$) the following effective action after integrating the hydrodynamical field $b_\mu$:
\begin{equation}
    S[A] = -\half  \left(\frac e {2\pi}\right)^{-2} \frac 1 {\rho_L^2 +\rho_H^2}\int\frac{d^3p }{(2\pi)^3} \, \frac 1 {\omega^2}\left[ \left(\frac e {2\pi}\right)^{2} i\omega\rho_L  \vec i\omega \vec A\cdot\vec E + \frac {k} {2\pi} i\omega  \epsilon^{ij} E_i i\omega A_j \right]\, ,
\end{equation}
giving the conductivity tensor,
\begin{equation}\label{condten}
\bar{\sigma} =  \frac 1 {\rho^2_L +\rho^2_H}\begin{pmatrix}
\rho_L & -\rho_H \\
\rho_H & \rho_L
\end{pmatrix} \, .
\end{equation}
which is the inverse of the resistivity tensor given by \eqref{results}. 

Expanding in the small quantity $\rho_L/\rho_H \sim \alpha$, we have,
\begin{equation} \label{hallcond}
    \sigma_H = \frac {f(\alpha)} {\rho_H} =\frac 1 {\rho_H} \left( 1+\frac{\rho^2_L}{\rho^2_H} \right)^{-1} =  \frac 1 {\rho_H} \left( 1- \left(\frac \alpha k \right)^2+{\cal O}(\alpha^4)\right) \, ,
\end{equation}
and thus confirm that while $\rho_H $ is quantized even in the presence of electromagnetism, $\sigma_H$ is not. 

Although the above was a more elaborate way to derive the conductivities than just inverting the resistivity tensor obtained in the main text {\it Hall conductance} part, it provided a check of the propagator in \eqref{propagator}, which we now use to calculate the fractional charge and statistics for the quasiparticles.

The part of the Lagrangian that is relevant for extracting the fractional charge is, 
\begin{align} \label{fraccharge}
    -\frac e {2\pi} \int\frac{d^3p }{(2\pi)^3} \,j^\mu G_{\mu\nu} \epsilon^{\nu\alpha\beta} \partial_\alpha A_\beta &= \frac e {2\pi} \left(\frac {2\pi} e\right)^4 \frac 1 {\rho^2_L +\rho^2_H} \frac k {2\pi} \int\frac{d^3p }{(2\pi)^3}\frac 1 {p^2} \left(\epsilon^{\mu\alpha\nu}p_\alpha A_\nu \epsilon_{\mu\beta\sigma}p^\beta j^\sigma j^\mu \right) \nonumber \\
    &= \frac e {2\pi} \int\frac{d^3p }{(2\pi)^3} f(\alpha) \frac e k  A_\mu j^\mu  \, ,
\end{align}
where we made a partial integration and dropped a pice that i zero in the $\vec q \rightarrow 0$ limit. 

In  the current-current interaction we drop the part $\sim j^\mu j_\mu$ and keep the part relevant for the fractional statistics 
\begin{equation} \label{potential}
 \half  \left(\frac {2\pi} e\right)^4  \frac 1 {\rho^2_L +\rho^2_H} \frac k{2\pi} \int\frac{d^3p }{(2\pi)^3} \frac {-i} {p^2} j^\mu \epsilon_{\mu\beta\nu}(-ip^\beta) j^\nu = \int\frac{d^3p }{(2\pi)^3}\, f(\alpha)\frac \pi k  \epsilon^{\mu\beta\nu} j_\mu  \frac {ip_\beta} {p^2} j_\nu \, .
\end{equation}
 From \eqref{fraccharge} and \eqref{potential} we see that both the original fractional charge $e^\star = e/k$, and the  statistical angle $\theta_s = \pi/k$ are renormalized by the same factor $f(\alpha)$ as the Hall conductivity.

 The above results in the $\tilde\sigma\ll\omega\rightarrow 0$ limit crucially depend on the fact that the 3d photons are massless. If the QH sample is encased a in cavity, the outcome is radically different. In \oncite{cardoso2026cavity} the same hydrodynamic description as in this paper is used to calculate $\sigma_H$ and $\sigma_{xx}$ for a QH liquid coupled to a single gapped 
 photon polarized in the $y$-direction. That situation is special since $\sigma_{yy}$ vanishes so both $\sigma_{H}$ and $\rho_H$ are quantized at their canonical values. However, it is easy to show that when one includes a second mode polarized in the $x$-direction, only $\rho_{H}$ is robust against the coupling to the cavity photons, in full accordance with the results in this paper. In contrast, in \oncite{yang2026quantum}, which contains a similar study in a chiral cavity, $\rho_H$ is modified, which is not surprising given that the electromagnetic environment breaks parity.

\noindent
{\bf IV. Derivation of $\sigma_H$ from the microscopic GLCS theory} 

Here we derive the electromagnetic response of a Hall liquid directly from the microscopic Ginzburg-Landau-Chern-Simons theory.
The  starting point is the microscopic theory of bosonized fermions of mass $m$ in a strong magnetic field \cite{zhk,zhangrev},
\begin{equation}  \label{bosact}
S_b = \int d^3x\,  \left[\phi^\star (i\partial_0  - a_0 +e A_0)\phi - \frac 1 {2m} \phi^\star\left(\frac 1 i\vec\nabla \  +\vec a -e \vec A\right)^2 \phi \\
 +  \frac 1 {2\pi k}\epsilon^{ij} a_0 \partial_i a_j  \right]
\end{equation}
where $\vec a$ is a transverse statistical gauge field, and the background potential $\vec A$ includes both the static background field, $B$ and a weak electric probe field. The last term $\sim 1/2\pi k$ is a  Chern-Simons action in Coulomb gauge.  one also includes an interaction term $V[\rho]$ which is a, generally nonlocal, functional of the density $\rho = \phi^\star \phi$.  Here such a term is not put by hand, but, as shown below, will be generated by an explicit coupling to a fluctuating electromagnetic field. 

We parametrize the boson field $\phi$ as 
\begin{align}\label{decomp}
\phi=\sqrt{\rho({x})}e^{i\theta({x})}\xi(x) \, ,
\end{align}
where  $\theta$ is a smooth phase, and $\xi $ is a singular phase factor that describes  a collection of  vortex singularities. Substituting into \eqref{bosact} and neglecting a total time derivative gives,
\begin{align}  \label{bosact2}
S_b = \int d^3x\,  &\left[\rho ( - a_0-\partial_0\theta +e A_0+i\xi^\star\partial_0\xi) -\frac \rho {2m} \left(\vec a  +\vec\nabla\theta -e \vec A-i\xi^\star\vec\nabla\xi \right)^2 \right.  \\
&\left.   +\frac \rho {8m}\vec\nabla \operatorname{ln}(\rho)   \cdot \vec\nabla \operatorname{ln}(\rho)
+  \frac 1 {2\pi k}\epsilon^{ij} a_0 \partial_i a_j  \right] 
\, . \nonumber
\end{align}
We now go to a unitary gauge by the regular gauge transformation,
\begin{equation}
    a_\mu \rightarrow a_\mu  - \partial_\mu\theta \, ,
\end{equation}
thus introducing a longitudinal component of the gauge field $a$, which then becomes dynamical, since the resulting covariant Chern-Simons action $\int d^3x\,\frac 1 {4\pi k}\epsilon^{\mu\nu\lambda}a_\mu\partial_\nu a_\lambda $ will provide the symplectic term $\frac 1 {2k\pi}a_x\partial_0 a_y$. With this, and neglecting the quasiparticles, we have,
\begin{align}  \label{bosact3}
S_b = \int d^3x\,  &\left[\rho ( - a_0 +e A_0+ eQ_0) -\frac \rho {2m} \left(\vec a   -e \vec A -e\vec Q \right)^2 \right.  \\
&\left.   +\frac \rho {8m}\vec\nabla \operatorname{ln}(\rho)   \cdot \vec\nabla \operatorname{ln}(\rho)
+  \frac 1 {4\pi k}\epsilon^{\mu\nu\lambda}a_\mu\partial_\nu a_\lambda   \right] 
  \, , \nonumber
\end{align}
where we introduced a fluctuating electromagnetic vector potential $Q_\mu$. 

As in the main text, we neglect terms $\vec\nabla (\rho) $,  integrate over $a_0$ to get the constraint $2\pi k \rho=B+\delta b$, where $\delta b = \varepsilon^{ij}\partial_i\delta a_j$, and  assume that $\rho = \bar\rho +\delta\rho$, where $2\pi \bar\rho = B$, and $\delta \rho$ is a small fluctuation away from the static homogeneous background density $\bar\rho$.
Setting $\vec a = e\vec A +\delta \vec a$ we are left with 
\begin{align}  \label{bosact4}
S_b = \int d^3x\,  &\left[\rho (  +e A_0 + eQ_0) -\frac \rho {2m} \delta\vec a \cdot  \delta\vec a
-  \frac 1 {4\pi k}\varepsilon^{ij}(eA +eQ + \delta a)_i\partial_0 (eA+eQ +\delta a)_j   \right] 
  \, , \nonumber
\end{align}
Neglecting the fluctuations and the dynamical gauge field we have $S_b = \frac {e^2} {4\pi k}\int d^3x \varepsilon^{ij}A_i E_j$ which gives $\sigma_H = \frac {e^2} {2\pi k}$. Since 
$$ \frac \rho {2m} = \frac 1 {4\pi k} \frac B m = \half \frac {\omega_c} {2\pi k}$$
integrating the massive field $\delta \vec a$ gives a term $\sim \omega/\omega_c$ which can be neglected.

In Coulomb gauge, $\vec\nabla\cdot\vec Q=0$,  the Maxwell action (with $\varepsilon_0 = 1$) becomes,  
\begin{align} 
  {\cal S}_{M} &=   \int\frac {d^4p }{(2\pi)^4}   \left(\frac 1 2\varepsilon (\vec k ,\omega) \vec E(-p)\cdot \vec E(p)  -\half \vec B(-p) \cdot \vec B(p)\right)  \\ 
   &=\frac 1 2 \int \frac{d^4p} {(2\pi)^4}\left[ \vec Q(-p)(\omega^2\varepsilon(\vec p,\omega)-q^2 -p_z^2  )\cdot \vec Q (p) +
 Q_0(-p)\, \varepsilon(\vec  p,\omega)p^2 Q_0 (p) \right]
\end{align}
where we again used the notation  $p =(\vec p, \omega) =(q_i,p_z,\omega)$, and  $q=|\vec q|$, with $\vec q$  a two-vector  in the plane of the QH liquid.  Currents and charges confined to the $(x,y)$ plane, couple only to the $p_z$ independent gauge fields, so as in the main text we can integrate $p_z$ to get the effective (2+1)D action,
\begin{align} \label{genact2}
  {\cal S}_{M} =\frac 1 2 \int \frac{d^3p} {(2\pi)^3} \vec Q(-p)2\sqrt{\omega^2\varepsilon(\vec p,\omega)-q^2 }\cdot \vec Q (p)  - V[\rho]  
\end{align}
where $  - V[\rho]  $ is the screened instantaneous Coulomb potential which is now recovered   from the $Q_0$ term.

To integrate $\vec Q$ we combine the kernel in \eqref{genact2} with the relevant terms in \eqref{bosact4} 
\begin{align} \label{genact3}
  {\cal S}_{\vec Q} =\frac 1 2 \int \frac{d^3p} {(2\pi)^3}  Q_i(-p)\left[2\sqrt{\omega^2\varepsilon(\vec p,\omega)-q^2 }\delta^{ij} + \frac {i\omega e^2}{2\pi k}\varepsilon^{ij}\right]  Q _j(p)- \frac {e^2} {2\pi k}\varepsilon^{ij}Q _i(-p) i\omega A_j (p)
\end{align}
where $\vec E =i\omega \vec A$ is the electric field. To extract the low frequency conductivities, we take $q=0$  and recall the relations, 
\begin{equation} \label{expr}
   \rho_H = k \frac{2\pi}{e^2} = kR_K \quad\quad ;\quad\quad \rho_L =   \half    \frac { \omega } {\sqrt{\omega^2 + i\tilde\sigma \omega  }} =     \half \frac { \omega } {\sqrt{\omega^2 + i\tilde\sigma \omega  }} Z_0\, .
\end{equation}
to get 
\begin{align} \label{low}
  {\cal S}_{\vec Q} =\frac 1 2 \int \frac{d^3p} {(2\pi)^3}  Q_i(-p)\left[\frac {i\omega}{\rho_L}\delta^{ij} + \frac {i\omega }{\rho_H}\varepsilon^{ij}\right]  Q _j(p)- \frac 1 {\rho_H}\varepsilon^{ij}A _i(-p) i\omega Q_j (p)
\end{align}
where $\sigma_0 = 1/\rho_H$. Integrating $\vec Q$ we get a nonzero $\sigma_L$ as well as a correction to the zeroth order $\sigma_H = \sigma_0$, the  correction to response action is,
\begin{align} \label{low}
  \Delta{\cal S}[{\vec A}] =\frac 1 2 \int \frac{d^3p} {(2\pi)^3}\frac {\rho_L^2}{\rho_L^2 +\rho_H^2} A_i(-\omega)\left[\frac 1{\rho_L}\delta^{ij} + \frac 1 { \rho_H}\varepsilon^{ij}\right]i\omega A_j(\omega)
\end{align}
adding the zeroth order contribution ${\cal L}_0[{\vec A}] = \frac {e^2}{ 2\pi k} A_i(-\omega) i\omega \varepsilon^{ij} A_j(\omega) $, and using \eqref{expr} the full conductivity tensor becomes,  
\begin{equation}\label{condten}
\bar{\sigma} =  \frac 1 {\rho^2_L +\rho^2_H}\begin{pmatrix}
\rho_L & -\rho_H \\
\rho_H & \rho_L
\end{pmatrix} \, .
\end{equation}
in accordance with the previous result obtained from the hydrodynamic Wen-Zee action.

\end{document}